\documentclass{ws-ijmpa}
\usepackage{graphicx}
\usepackage[super,compress]{cite}
\usepackage{epstopdf}
\usepackage{hyperref}
\usepackage{float}

\begin{document}

\def\be{\begin{equation}}
\def\ee{\end{equation}}
\def\ba{\begin{array}}
\def\ea{\end{array}}
\def\bea{\begin{eqnarray}}
\def\eea{\end{eqnarray}}
\def\bc{\begin{center}}
\def\ec{\end{center}}



\title{Electromagnetic and gravitational form factors in simulated QED and Yukawa model}
\author{Narinder Kumar and Harleen Dahiya}
\address{Department of Physics\\
Dr. B. R. Ambedkar National Institute of Technology\\
         Jalandhar-144011, India}
\maketitle
\begin{abstract}
The light-cone Fock state representation of composite systems has number of remarkable properties and for systems such as hadrons they have exact representation for angular momentum, energy momentum tensor. We investigate the electromagnetic and gravitational form factors with zero momentum transfer in QED and Yukawa theory.  To improve the convergence near the end points of $x$ qualitatively as well as to check the consistency of the model, we differentiate the wavefunction w.r.t. bound state mass. We test the behaviour of the anomalous gravitomagnetic moment, which follows directly from the Lorentz boost properties of the light-cone Fock representation, for the simulated model as well as the Yukawa model. We also discuss the Pauli form factor obtained from the spin-flip matrix element.
\end{abstract}
\section{Introduction}
One of most outstanding problem in Quantum Chromodynamics (QCD) is to study the internal structure of hadrons i.e. to determine the spectrum and structure of hadrons in terms of their quark and gluon degrees of freedom.  The light-cone Fock state wavefunction (LCWF) ${\psi_{n/H}(x_i , \vec{k_{\perp i}}, \lambda_i)}$\cite{mod,mod1,mod2} has number of remarkable features. The set of LCWFs provide a frame-independent, quantum-mechanical description of hadrons at the amplitude level which are capable of encoding multi-quark and gluon momentum, helicity and flavor correlations in the form of universal process independent
hadron wavefunctions. One can also construct the invariant mass operator $H_{LC}= P^+ P^- - P_\perp^2$ and light-cone time operator $P^-=P^0-P^z$  in the light-cone gauge from the QCD Lagrangian\cite{rev,wf,wf1}.
The coordinates of the light-cone Fock wavefunctions $\psi_{n/H}(x_i,\vec
k_{\perp i},\lambda_i)$ are the light-cone momentum fractions $x_i = k^+_i/P^+$, ${\vec
k_{\perp i}}$ represent the relative momentum coordinates of the QCD
constituents and $\lambda_i$ label the spin projections of the quarks in the $z$-direction.  The physical transverse momenta are represented as ${\vec p_{\perp i}}
= x_i {\vec P_\perp} + {\vec k_{\perp i}}.$ The physical gluon
polarization vectors
$\epsilon^\mu(k,\ \lambda = \pm 1)$ are specified in light-cone
gauge by the conditions $k \cdot \epsilon = 0,\ \eta \cdot \epsilon =
\epsilon^+ = 0.$
Light-cone Fock state wavefunction satisfies the conservation of the projection of angular momentum: $J^z= \sum_{i=1}^{n} S_i^z + \sum_{j=1}^{n-1} l_z^j$, where sum over the spin $S_i^z$ represents the contribution of spin from $n$ Fock state constituents and the sum over the angular momentum is derived from $n-1$ relative momenta, here angular momentum is represented as $l_j^z=(-k_j^1 \frac{\partial}{\partial k_j^2}- k_j^2 \frac{\partial}{\partial k_j^1})$. It excludes the contribution to the orbital angular momentum due to the motion of the center of mass, which is not an intrinsic property of the hadron. The light-cone wavefunctions also specify the distribution of the spectator particles in the final state\cite{mod1} which could be measured in the proton fragmentation region in Deep Inelastic Scattering (DIS) experiments. One can compute the transversity distribution and moment of helicity, measurable in the DIS experiments, from the light-cone wavefunctions.

We present a simple self consistent model of an effective composite spin-$\frac{1}{2}$ system based on the quantum fluctuation of the electron in Quantum Electrodynamics (QED). The wavefunctions generated by the radiative corrections to the electron in QED provide an ideal system for understanding the spin and angular momentum decomposition of relativistic systems. The LCWFs of an electron can be evaluated in QED perturbation theory\cite{mod,mod1}. We represent a spin-$\frac{1}{2}$ system as a composite of spin-$\frac{1}{2}$  fermion and spin-1 vector boson  and spin-0 scalar boson with arbitrary masses\cite{wf,wf1}. We consider the composite
system consisting of fermion state with a mass $M$ and a vector constituent with respective masses $m$ and $\lambda$.
From the diagonal overlap of the light-cone wavefunctions, space like electromagnetic, electroweak or gravitational form factor of a composite or elementary system can be evaluated. Gravitational form factors for composite hadrons have also been studied in light front holography\cite{hadrons,holography,holography1,LFDandADS} which is a remarkable feature of AdS/QCD. Form factors have been measured in many experiments\cite{jeffer,jeffer1,babar,compass}.  Form factors with strangeness contribution have been also measured explicitly\cite{sample,a4,happex}.
Gravitational form factors have also been studied in chiral quark model\cite{gra,gra1}.

In the present work, we study the spin-flip matter form factor $B(q^2)$ of energy-momentum tensor for a spin-$\frac{1}{2}$ composite system. The spin-flip matter form factor receives contribution from fermion and boson constituents. In Ref.\cite{wf1} it was proved that anomalous gravitomagnetic moment coupling $B(0)$ to gravity vanish for any composite system. Classically, this result was derived from the equivalence principle\cite{Okun,equp} and from the conservation of the energy-momentum tensor\cite{Kob70}. Further, in Ref.\cite{imp} it has also been shown that differentiating the wavefunction w.r.t bound state mass $M^2$ improves the behavior of wavefunctions near the end points of $x$ . In this context, we intend to check the consistency of results i.e. whether the contribution of fermion and boson constituents to $B(q^2)$, in the case of simulated model, vanishes at zero momentum transfer or not. We calculate the fermion and boson contributions to the spin-flip matter form factor in QED and Yukawa theory in simulated model. For $q^2\ne0$, $B(q^2)$ does not vanish. However, this model will provide a check whether the contributions to the spin-flip matter form factor\cite{lorentz,lorentz1,lorentz2,lorentz3} vanishes at $q^2=0$ or not due to the Lorentz boost properties. The LCWFs have a number of remarkable properties. The matrix elements of space-like local operators such as currents, angular momentum and the energy-momentum tensor for the coupling of photons, gravitons, and the moments of deep inelastic structure functions  can be expressed as overlaps of LCWFs with the same number of Fock constituents and have exact Lorentz invariant representations. We have shown that after summing the  graviton couplings to each of the n- constituents,  contribution to B(0) vanishes for each Fock component due to the Lorentz boost properties of the light-cone Fock representation. The anomalous magnetic moment can be obtained from the Pauli form factor. However, the magnetic moment of the composite system can be obtained from both Pauli and Dirac form factor in the limit of zero momentum transfer. In this context, we have also studied the Pauli form factor $F_2(q^2)$ obtained from the helicity-flip vector current matrix elements of the $J^+$ current\cite{mod} in the simulated model.

The plan of the paper is as follows. To make the manuscript readable
as well as to facilitate discussion, in Sec 2 we present
some of the essentials of electromagnetic and gravitational form factors. In Sec 3,  the results for simulated QED model have been presented.
Sec 4 presents the results for simulated Yukawa model. Sec 5 comprises the summary and the conclusions.

\section{Electromagnetic and Gravitational Form Factors}
The matrix elements of local operators like energy momentum tensor, electromagnetic tensor and moment of structure functions have exact representation in light-cone Fock state wavefunctions of bound states such as hadrons. Given the local operators for the energy momentum tensor $T^{\mu\nu}(x)$ and the angular momentum tensor $M^{\mu\nu\lambda} (x)$, we can directly compute momentum fraction, gravitomagnetic moments and the form factors appearing in the coupling of gravitons to composite systems. The gravitomagnetic form factors of the energy momentum tensor $A(q^2)$ and $B(q^2)$ for a spin-$\frac{1}{2}$ composite are defined as\cite{wf1}
\begin{eqnarray}
      \langle P'| T^{\mu\nu} (0)|P \rangle
       &=& \bar u(P')\, \Big[\, A(q^2)
       \gamma^{(\mu} \bar P^{\nu)} +
   B(q^2){i\over 2M} \bar P^{(\mu} \sigma^{\nu)\alpha}
q_\alpha \nonumber \\
   &&\qquad\qquad +  C(q^2){1\over M}(q^\mu q^\nu - g^{\mu\nu}q^2)
    \, \Big]\, u(P) \ ,
\label{Ji12}
\end{eqnarray}
where $\bar P^\mu={1\over 2}(P'+P)^\mu$, $q^\mu = (P'-P)^\mu$,
$a^{(\mu}b^{\nu)}={1\over 2}(a^\mu b^\nu +a^\nu b^\mu)$,
and $u(P)$ is the spinor.
By calculating only the non-interacting part, from the ++ component of Eq. (1), we get
\begin{equation}
\langle{P+q,\uparrow|\frac{T^{++}(0)}{2(P^+)^2}
|P,\uparrow}\rangle =A(q^2)\ ,
\label{eBD1}
\end{equation}
\begin{equation}
\langle{P+q,\uparrow|\frac{T^{++}(0)}{2(P^+)^2}|P,\downarrow}\rangle=
-(q^1-{\mathrm i} q^2){B(q^2)\over 2M}\ .
\label{eBD2}
\end{equation}
The $A(q^2)$ and $B(q^2)$ form factors\cite{lorentz,lorentz1} in
Eqs. (\ref{eBD1}) and (\ref{eBD2})
are very similar to the Dirac and Pauli form
factors defined as
\be
\langle P'|J^\mu(0)|P  \rangle= \bar{u}(P')[F_1(q^2) \gamma^\mu + F_2(q^2) \frac{i}{2 M} \sigma^{\mu \alpha} q_\alpha]u(P).
\ee
The Dirac and Pauli form factors can further be derived from the helicity non-flip and helicity flip vector current matrix elements of the $J^+$ current as follows
\bea
\langle P + q ,\uparrow |\frac{J^+(0)}{2 P^+}|P ,\uparrow \rangle &=& F_1(q^2) , \nonumber\\
\langle P + q ,\uparrow |\frac{J^+(0)}{2 P^+}|P ,\downarrow \rangle &=& - (q^1-i q^2) \frac{F_2(q^2)}{2M} .
\label{diracnpauli}
\eea
\section{Simulated model calculations in QED model}
After discussing the general aspects of the electromagnetic and gravitational form factors, we now present the essentials of the QED model. The two-particle Fock state for an electron with $J^z=+\frac{1}{2}$ has four possible spin combinations:
\begin{eqnarray}
&&\left|\Psi^{\uparrow}_{\rm two \ particle}(P^+, \vec P_\perp = \vec
0_\perp)\right>
\ =\
\int\frac{{\mathrm d}^2 {\vec k}_{\perp} {\mathrm d} x }{{\sqrt{x(1-x)}}16
\pi^3}
\label{fockup}\\
&\times&
\Big[ \
\psi^{\uparrow}_{+\frac{1}{2}\, +1}(x,{\vec k}_{\perp})\,
\left| +\frac{1}{2}\, +1\, ;\,\, xP^+\, ,\,\, {\vec k}_{\perp}\right>
+\psi^{\uparrow}_{+\frac{1}{2}\, -1}(x,{\vec k}_{\perp})\,
\left| +\frac{1}{2}\, -1\, ;\,\, xP^+\, ,\,\, {\vec k}_{\perp}\right>
\nonumber\\
&&\ \ \ +\psi^{\uparrow}_{-\frac{1}{2}\, +1} (x,{\vec k}_{\perp})\,
\left| -\frac{1}{2}\, +1\, ;\,\, xP^+\, ,\,\, {\vec k}_{\perp}\right>
+\psi^{\uparrow}_{-\frac{1}{2}\, -1} (x,{\vec k}_{\perp})\,
\left| -\frac{1}{2}\, -1\, ;\,\, xP^+\, ,\,\, {\vec k}_{\perp}\right>\ \Big].
\
\nonumber
\end{eqnarray}
The two-particle wavefunctions for spin-up electron can be expressed as\cite{mod2,wf,wf1,imp,stan}
\begin{equation}
\left
\{ \begin{array}{l}
\psi^{\uparrow}_{+\frac{1}{2}\, +1} (x,{\vec k}_{\perp})=-{\sqrt{2}}
\ \frac{-k^1+{i} k^2}{x(1-x)}\,
\varphi \ ,\\
\psi^{\uparrow}_{+\frac{1}{2}\, -1} (x,{\vec k}_{\perp})=-{\sqrt{2}}
\ \frac{k^1+{i} k^2}{1-x }\,
\varphi \ ,\\
\psi^{\uparrow}_{-\frac{1}{2}\, +1} (x,{\vec k}_{\perp})=-{\sqrt{2}}
\ (M-{m\over x})\,
\varphi \ ,\\
\psi^{\uparrow}_{-\frac{1}{2}\, -1} (x,{\vec k}_{\perp})=0\ ,
\end{array}
\right.
\label{spinup}
\end{equation}
where \be
\varphi (x,{\vec k}_{\perp}) = \frac{e}{\sqrt{1-x}}\
\frac{1}{M^2-{{\vec k}_{\perp}^2+m^2 \over x}
-{{\vec k}_{\perp}^2+\lambda^2 \over 1-x}}\, .
\label{varphi}
\ee
%
We work in a generalized form of QED
by assigning a mass $M$ to the external electrons and a different
mass $m$ to the internal electron lines and a mass $\lambda$ to the
internal photon lines.

The wavefunction for an electron with negative
helicity can similarly be obtained and can be expressed as
\begin{eqnarray}
&&\left|\Psi^{\downarrow}_{\rm two \ particle}(P^+, \vec P_\perp =
\vec 0_\perp)\right>
\ =\
\int\frac{{\mathrm d}^2 {\vec k}_{\perp} {\mathrm d} x }{{\sqrt{x(1-x)}}16
\pi^3}
\label{fockdown}\\
&\times&
\Big[\
\psi^{\downarrow}_{+\frac{1}{2}\, +1}(x,{\vec k}_{\perp})\,
\left| +\frac{1}{2}\, +1\, ;\,\, xP^+\, ,\,\, {\vec k}_{\perp}\right>
+\psi^{\downarrow}_{+\frac{1}{2}\, -1}(x,{\vec k}_{\perp})\,
\left| +\frac{1}{2}\, -1\, ;\,\, xP^+\, ,\,\, {\vec k}_{\perp}\right>
\nonumber\\
&&\ \ \ +\psi^{\downarrow}_{-\frac{1}{2}\, +1}(x,{\vec k}_{\perp})\,
\left| -\frac{1}{2}\, +1\, ;\,\, xP^+\, ,\,\, {\vec k}_{\perp}\right>
+\psi^{\downarrow}_{-\frac{1}{2}\, -1}(x,{\vec k}_{\perp})\,
\left| -\frac{1}{2}\, -1\, ;\,\, xP^+\, ,\,\, {\vec k}_{\perp}\right>\ \Big]
\ ,
\nonumber
\end{eqnarray}
where
\begin{equation}
\left
\{ \begin{array}{l}
\psi^{\downarrow}_{+\frac{1}{2}\, +1} (x,{\vec k}_{\perp})=0\ ,\\
\psi^{\downarrow}_{+\frac{1}{2}\, -1} (x,{\vec k}_{\perp})=-{\sqrt{2}}
(M-{m\over x})\,
\varphi \ ,\\
\psi^{\downarrow}_{-\frac{1}{2}\, +1} (x,{\vec k}_{\perp})=-{\sqrt{2}}
\frac{(-k^1+{\mathrm i} k^2)}{1-x }\,
\varphi \ ,\\
\psi^{\downarrow}_{-\frac{1}{2}\, -1} (x,{\vec k}_{\perp})=-{\sqrt{2}}
\frac{(+k^1+{\mathrm i} k^2)}{x(1-x)}\,
\varphi \ .
\end{array}
\right.
\label{spindown}
\end{equation}
The coefficients of $\varphi$ in Eqs. (\ref{spinup})
and (\ref{spindown}) are the matrix elements of
$\frac{\overline{u}(k^+, k^-, {\vec k}_{\perp})}{{\sqrt{k^+}}}
\gamma \cdot \epsilon^{*}
\frac{u (P^+, P^-, {\vec P}_{\perp})}{{\sqrt{P^+}}}$
which are
the numerators of the wavefunctions corresponding to
each constituent spin $s^z$ configuration.

As discussed in Ref.\cite{imp}
a differentiation of the QED LFWFs with respect to $M^2$ improves the
convergence of the wavefunctions at the end points: $x=0,1$
as well as the  $k^2_\perp$ behaviour,
thus simulating a bound state valence wavefunction.
In other words, we take
\begin{equation}
\varphi' (x,{\vec k}_{\perp}) = \mid {\partial \varphi (x,{\vec
k}_{\perp})\over
\partial M^2} \mid =  \frac{e}{\sqrt{1-x}}\
\frac{1}{\Big (M^2-{{\vec k}_{\perp}^2+m^2 \over x}
-{{\vec k}_{\perp}^2+\lambda^2 \over 1-x}\Big )^2}.
\label{simulated}
\end{equation}

The fermion and boson contributions to the spin-flip matter form factor
can now be expressed in terms of the two-particle wavefunctions giving
\begin{eqnarray}
B_{\rm f}(q^2)&=&
{-2M\over (q^1-{\mathrm i}q^2)}
\left<\Psi^{\uparrow}(P^+,{\vec P_\perp}={\vec q_\perp})
\right|\frac{T^{++}_{\rm f}(0)}{2(P^+)^2}
\left|\Psi^{\downarrow}(P^+,{\vec P_\perp}={\vec
0_\perp})\right>\nonumber\\
&=&{-2M\over (q^1-{\mathrm i}q^2)}\int\frac{{\mathrm d}^2 {\vec k}_{\perp} {\mathrm d} x }{16 \pi^3}\ x
\Big[\psi^{\uparrow\ *}_{+\frac{1}{2}\, -1}(x,{\vec
k'}_{\perp})
\psi^{\downarrow}_{+\frac{1}{2}\, -1}(x,{\vec k}_{\perp})
+\nonumber\\
&& \psi^{\uparrow\ *}_{-\frac{1}{2}\, +1}(x,{\vec k'}_{\perp})
\psi^{\downarrow}_{-\frac{1}{2}\, +1}(x,{\vec k}_{\perp})
\Big],
\label{bf}
\end{eqnarray}
\begin{eqnarray}
B_{\rm b}(q^2)&=&
{-2M\over (q^1-{\mathrm i}q^2)}
\left<\Psi^{\uparrow}(P^+,{\vec P_\perp}={\vec q_\perp})
\right|\frac{T^{++}_{\rm b}(0)}{2(P^+)^2}
\left|\Psi^{\downarrow}(P^+,{\vec P_\perp}={\vec
0_\perp})\right>
\nonumber\\
&=&{-2M\over (q^1-{\mathrm i}q^2)}
\int\frac{{\mathrm d}^2 {\vec k}_{\perp} {\mathrm d} x }{16 \pi^3}\ (1-x)
\Big[\psi^{\uparrow\ *}_{+\frac{1}{2}\, -1}(x,{\vec
k''}_{\perp})
\psi^{\downarrow}_{+\frac{1}{2}\, -1}(x,{\vec k}_{\perp})+\nonumber\\
&& \psi^{\uparrow\ *}_{-\frac{1}{2}\, +1}(x,{\vec k''}_{\perp})
\psi^{\downarrow}_{-\frac{1}{2}\, +1}(x,{\vec k}_{\perp})
\Big],
\label{Bb}
\end{eqnarray}
where
\begin{equation}
\vec{k}'_\perp=\vec{k}_\perp + (1-x) \vec{q}_\perp,
\end{equation}
and
\begin{equation}
\vec{k}''_\perp=\vec{k}_\perp - x \vec{q}_\perp.
\end{equation}
Now using this simulated wavefunction given in Eq. (\ref{simulated}) we get
\begin{eqnarray}
B_f(q^2)&=& - \frac{2 M}{q_\perp} \frac{x^4(1-x)^3}{L_1^2 L_2^2} \int \frac{d^2k_\perp dx}{16 \pi^3} x \Big[\frac{2(k_\perp+ (1-x) q_\perp)}{1-x} \left(M- \frac{m}{x}\right) - 2 \left(M- \frac{m}{x}\right) \frac{k_\perp}{1-x}\Big]\nonumber\\
&=&- 4 M  \frac{e^2}{16 \pi^3} \int dx \ d^2 k_\perp \frac{x^5 (1-x)^3 (M-\frac{m}{x})}{L_1^2 \ L_2^2} \nonumber\\
&=& - 4 M  \frac{e^2}{16 \pi^3} \int dx \ x^5 (1-x)^3 \left(M-\frac{m}{x}\right) I_1,
\label{bffinal}
\end{eqnarray}
and
\begin{eqnarray}
B_b(q^2)&=& - \frac{2 M}{q_\perp} \frac{x^4(1-x)^3}{L_1^2 L_2^2} \int \frac{d^2k_\perp dx}{16 \pi^3} (1-x) \Big[\frac{2(k_\perp -x q_\perp)}{1-x} \left(M- \frac{m}{x}\right) - 2 \left(M- \frac{m}{x}\right) \frac{k_\perp}{1-x}\Big]\nonumber\\
&=& 4 M  \frac{ e^2}{16 \pi^3}\int d^2k_\perp dx \frac{x^5 (1-x)^3 \left(M-\frac{m}{x}\right)}{L_3^2 \ L_4^2}\nonumber\\
&=& 4 M  \frac{e^2}{16 \pi^3} \int dx \ x^5 (1-x)^3 \left(M-\frac{m}{x}\right) I_2.
\label{Bbfinal}
\end{eqnarray}
Here
\bea
I_1 &=& \int {d^2 \vec{k}_\perp \over L_1^2 L_2^2}= \pi \int \frac{\alpha(1-\alpha)}{D^3} d\alpha \nonumber \,,\\
I_2 &=& \int {d^2 \vec{k}_\perp \over L_3^2 L_1^2}=  \pi \int_0^1 d \alpha {\alpha
(1-\alpha)\over D_1^3}, \
\label{integrals}
\eea
and
\bea
L_1 &=& k_\perp^2- M^2 x (1-x) + m^2 (1-x)+\lambda^2 x, \nonumber\\ \
L_2 &=& k_\perp^2 + (1-x)^2 q_\perp^2- 2 (1-x) k_\perp \cdot q_\perp - M^2 x (1-x) + m^2 (1-x)+\lambda^2 x, \nonumber\\ \
L_3 &=& k_\perp^2 - 2 x  k_\perp \cdot q_\perp  +q_\perp^2 x^2 -M^2 x (1-x) + m^2 (1-x)- \lambda^2 x, \nonumber\\ \
D &=& \alpha (1-\alpha) (1-x)^2 q_{\perp}^2  - M^2 x (1 - x) + m^2 (1 - x) + \lambda^2 x, \nonumber\\ \
D_1 &=& \alpha (1-\alpha) x^2 q_{\perp}^2- M^2 x  (1-x)+ m^2 (1 - x) + \lambda^2 x.
\label{denominators}
\eea
The Pauli form factor is obtained from the spin-flip matrix element of $J^+$ current. From Eqs. (\ref{diracnpauli}), (\ref{fockup}) and (\ref{fockdown}) we have
\bea
F_2(q^2)&=& \frac{-2 M}{q^1- i q^2} \left<\Psi^{\uparrow}(P^+,{\vec P_\perp}={\vec q_\perp})
|
\Psi^{\downarrow}(P^+,{\vec P_\perp}={\vec
0_\perp})\right>\nonumber\\
&=& \frac{-2 M}{q^1- i q^2} \int\frac{{\mathrm d}^2 {\vec k}_{\perp} {\mathrm d} x }{16 \pi^3}\ x
\Big[\psi^{\uparrow\ *}_{+\frac{1}{2}\, -1}(x,{\vec
k'}_{\perp})
\psi^{\downarrow}_{+\frac{1}{2}\, -1}(x,{\vec k}_{\perp})
+\nonumber\\
&& \psi^{\uparrow\ *}_{-\frac{1}{2}\, +1}(x,{\vec k'}_{\perp})
\psi^{\downarrow}_{-\frac{1}{2}\, +1}(x,{\vec k}_{\perp})
\Big]\nonumber\\
&=& -4 M \frac{e^2}{16 \pi^3} \int dx \left(M-\frac{m}{x}\right)x^4 (1-x)^3 I_1.
\eea

The total contribution to spin-flip matter form factor from fermion and boson constituents is given by
\bea
B(q^2)&=&B_f(q^2)+B_b(q^2)\nonumber\\
&=& 2 M  \frac{e^2}{16 \pi^3} \int dx \left(- x^5 (1-x)^3 \left(M-\frac{m}{x}\right) I_1+\right.\nonumber\\
&& \left.x^5 (1-x)^4 \left(M-\frac{m}{x}\right) I_2\right).
\label{Bfinal}
\eea
\begin{figure}
\minipage{0.42\textwidth}
    \includegraphics[width=5.8cm ,angle=360]{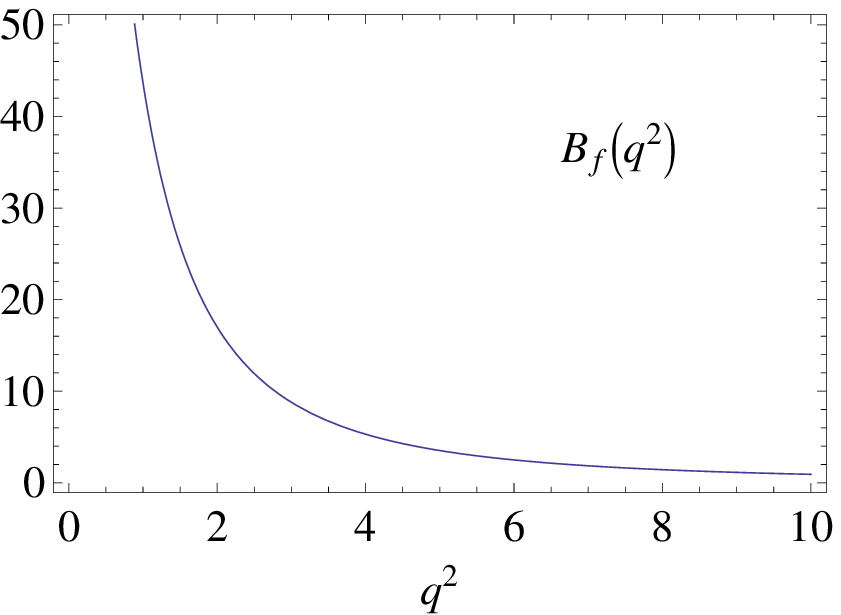}
  \endminipage\hfill
  \minipage{0.42\textwidth}
  \includegraphics[width=5.8cm ,angle=360]{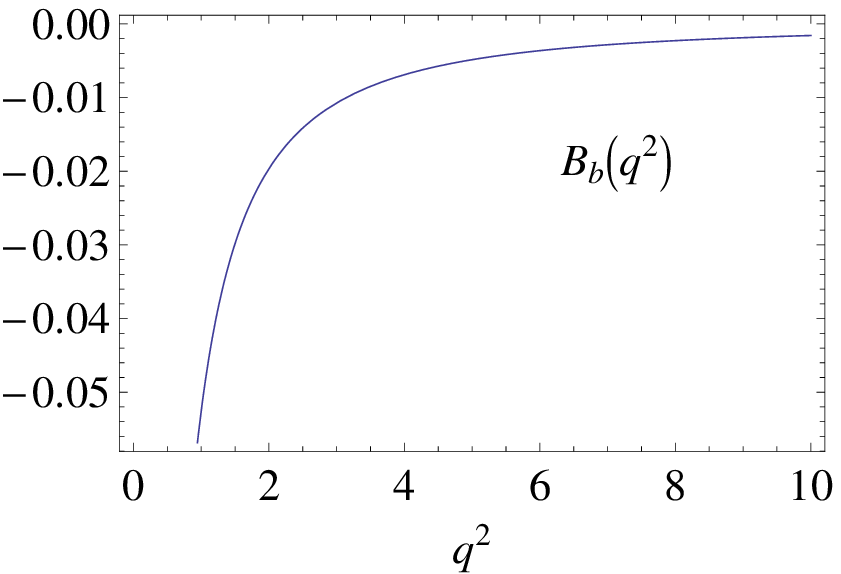}
\endminipage\hfill
\caption{Helicity-flip gravitational form factors for simulated model in QED.}
\label{qed}
\end{figure}
\begin{figure}
\minipage{0.42\textwidth}
\includegraphics[width=6cm ,angle=360]{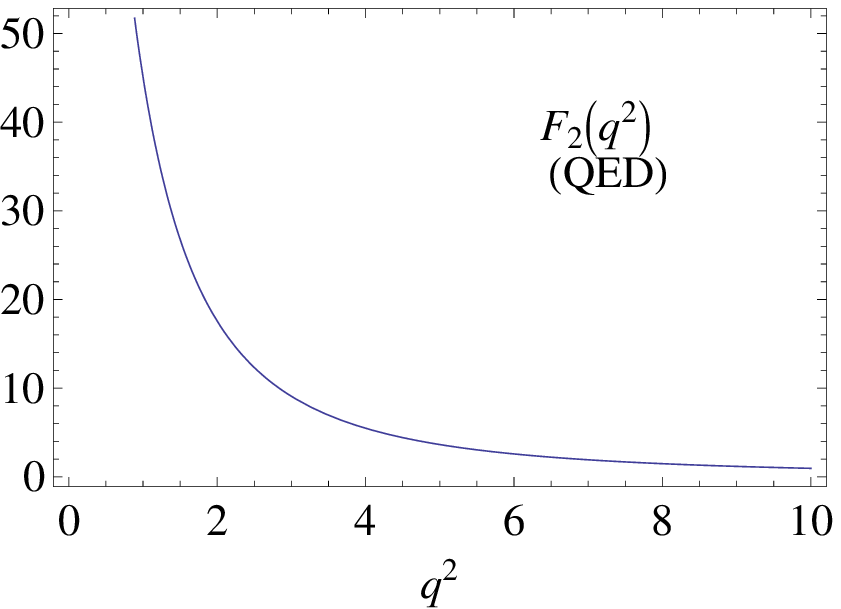}
\endminipage\hfill
\minipage{0.42\textwidth}
\includegraphics[width=6cm ,angle=360]{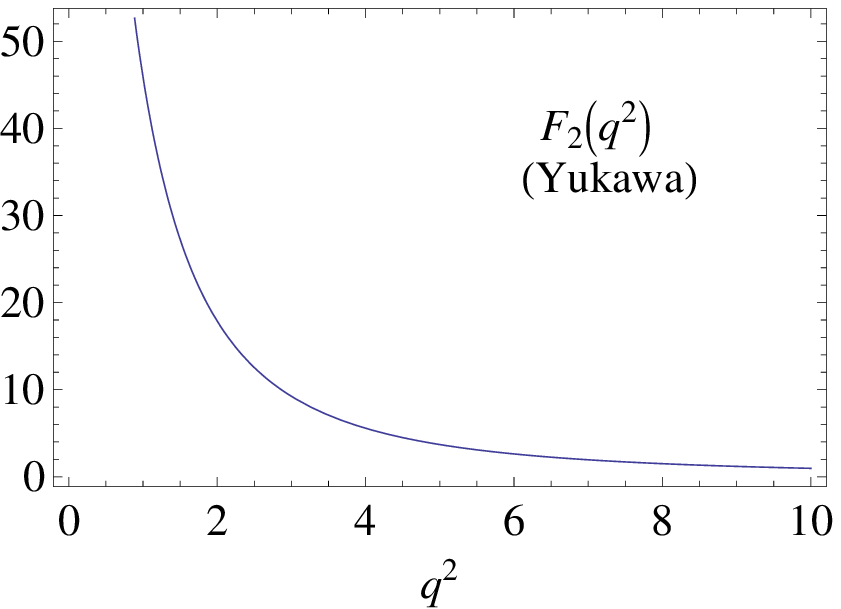}
\endminipage\hfill
\caption{Pauli form factor in QED model and Yukawa theory.}
\label{pauliformfactor}
\end{figure}
\begin{figure}
 \minipage{0.42\textwidth}
  \includegraphics[width=6cm,angle=360]{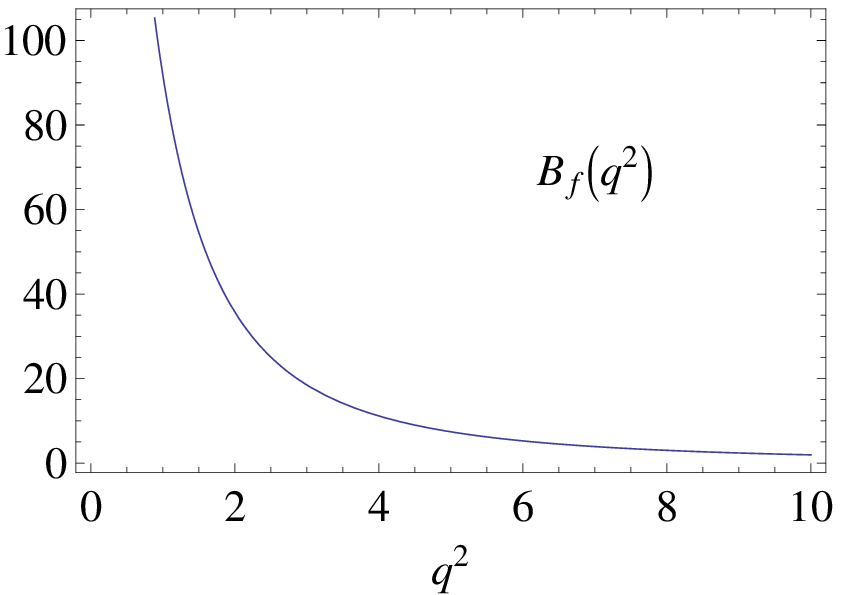}
  \endminipage\hfill
   \minipage{0.42\textwidth}
  \includegraphics[width=6cm,angle=360]{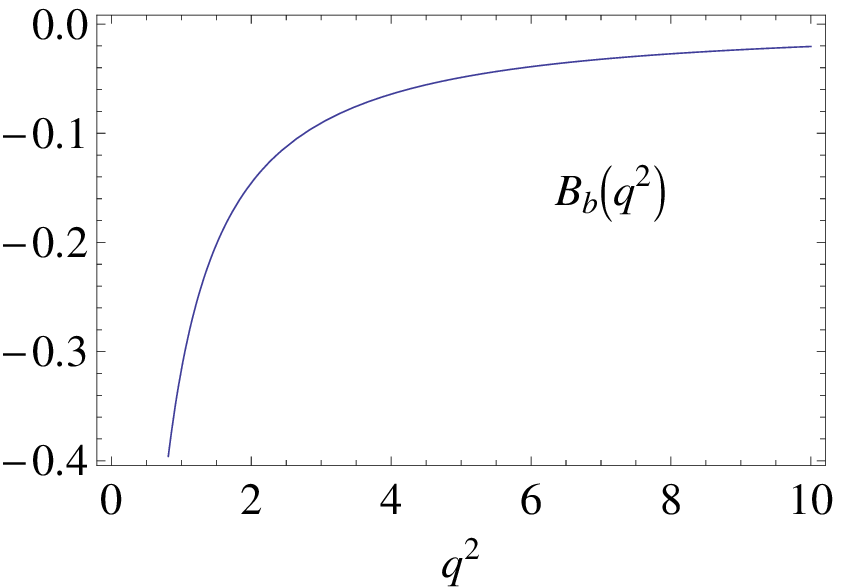}
  \endminipage\hfill
  \caption{Helicity-flip gravitational form factors for Yukawa theory.}
  \label{yukawa}
  \end{figure}

In order to check the behaviour of the fermion and boson constituent at zero momentum transfer we take $q_\perp^2 \rightarrow 0$ in the quantities which are dependent of $q_\perp^2$. It is found that at zero momentum transfer
\begin{eqnarray}
B(0)=B_f(0)+B_b(0)=0.
\end{eqnarray}
This result is in agreement with conservation of the energy momentum transfer and equivalence principle based classical arguments. For
numerical calculations we have taken $M=m=0.51$ MeV\cite{mod}. The helicity-flip boson form factor $B_b(q^2)$ of the graviton coupling to the boson constituent of the electron at one-loop order in QED comes out to be $B_b(0)=-226.27$ at $q_\perp^2 \rightarrow 0$. As expected, the helicity-flip fermion form factor $B_f(q^2)$ of the graviton coupling to the fermion constituent at one-loop order in QED comes out to be $B_f(0)=226.27$ at $q_\perp^2 \rightarrow 0$ leading to the cancellation of the graviton couplings to the boson and fermion constituent. The Pauli form factor for $q_\perp^2 \rightarrow 0$ comes out to be $F_2(0)=233.745$.
In Fig. \ref{qed} we have presented the results for the gravitational form factors as a function of $q^2$ in the simulated QED. From plot it is clear that $B_f(q^2)$ and $B_b(q^2)$ exhibit opposite behaviour which leads to the vanishing result for gravitomagnetic moment. In Fig. \ref{pauliformfactor}(a), we present the Pauli form factor in the QED model.

\section{Simulated Model Calculations in Yukawa theory}
In this section, we consider the composite system composed of a fermion and a neutral scaler based on the one-loop fluctuations.
The $J^z=+\frac{1}{2}$ two-particle Fock state is given by\cite{wf1}
\begin{eqnarray}
&&\left|\Psi^{\uparrow}_{\rm two \ particle}(P^+, \vec P_\perp = \vec
0_\perp)\right>
\label{sn1}\\
&=&
\int\frac{{\mathrm d}^2 {\vec k}_{\perp} {\mathrm d} x }{{\sqrt{x(1-x)}}16
\pi^3}
\Big[ \
\psi^{\uparrow}_{+\frac{1}{2}} (x,{\vec k}_{\perp})\,
\left| +\frac{1}{2}\, ;\,\, xP^+\, ,\,\, {\vec k}_{\perp} \right>
+\psi^{\uparrow}_{-\frac{1}{2}} (x,{\vec k}_{\perp})\,
\left| -\frac{1}{2}\, ;\,\, xP^+\, ,\,\, {\vec k}_{\perp} \right>\ \Big]\ ,
\nonumber
\end{eqnarray}

where
\begin{equation}
\left
\{ \begin{array}{l}
\psi^{\uparrow}_{+\frac{1}{2}} (x,{\vec k}_{\perp})=(M+\frac{m}{x})\,
\varphi \ ,\\
\psi^{\uparrow}_{-\frac{1}{2}} (x,{\vec k}_{\perp})=
-\frac{(+k^1+{\mathrm i} k^2)}{x }\,
\varphi \ .
\end{array}
\right.
\label{sn2}
\end{equation}
The scalar part of the wavefunction $\varphi$
is given by Eq. (\ref{varphi}). Here we have replace $e$ by $g$ in the numerator representing the Yukawa coupling.

The $J^z = - {1\over 2}$ two-particle Fock state is defined in a similar manner and is given by
\begin{eqnarray}
&&\left|\Psi^{\downarrow}_{\rm two \ particle}(P^+, \vec P_\perp =
\vec 0_\perp)\right>
\label{sn1a}\\
&=&
\int\frac{{\mathrm d}^2 {\vec k}_{\perp} {\mathrm d} x }{{\sqrt{x(1-x)}}16
\pi^3}
\Big[ \
\psi^{\downarrow}_{+\frac{1}{2}} (x,{\vec k}_{\perp})\,
\left| +\frac{1}{2}\, ;\,\, xP^+\, ,\,\, {\vec k}_{\perp} \right>
+\psi^{\downarrow}_{-\frac{1}{2}} (x,{\vec k}_{\perp})\,
\left| -\frac{1}{2}\, ;\,\, xP^+\, ,\,\, {\vec k}_{\perp} \right>\ \Big]\ ,
\nonumber
\end{eqnarray}
where
\begin{equation}
\left
\{ \begin{array}{l}
\psi^{\downarrow}_{+\frac{1}{2}} (x,{\vec k}_{\perp})=
\frac{(+k^1-{\mathrm i} k^2)}{x }\,
\varphi \ ,\\
\psi^{\downarrow}_{-\frac{1}{2}} (x,{\vec k}_{\perp})=(M+\frac{m}{x})\,
\varphi \ .
\end{array}
\right.
\label{sn2a}
\end{equation}
The framework of the Yukawa theory has been generalized in Eqs. (\ref{sn2}) and (\ref{sn2a}) where we represent the structure
of a composite fermion state with mass $M$ by a fermion and a scalar boson
constituent with respective masses $m$ and $\lambda$.

The fermion and boson contributions to the spin-flip matter form factor are expressed as
\begin{eqnarray}
B_{\rm f}(q^2)
&=&{-2M\over (q^1-{\mathrm i}q^2)}
\int\frac{{\mathrm d}^2 {\vec k}_{\perp} {\mathrm d} x }{16 \pi^3}\ x\
\Big[\psi^{\uparrow\ *}_{+\frac{1}{2}}(x,{\vec k'}_{\perp})
\psi^{\downarrow}_{+\frac{1}{2}}(x,{\vec k}_{\perp})
+\psi^{\uparrow\ *}_{-\frac{1}{2}}(x,{\vec k'}_{\perp})
\psi^{\downarrow}_{-\frac{1}{2}}(x,{\vec k}_{\perp})
\Big]\nonumber\\
&=& - \frac{2 M}{q_\perp} \frac{g^2}{16 \pi^3} \int\frac{d^2 k_\perp dx}{16 \pi^3} \frac{x^5 (1-x)^3}{L_1^2 L_2^2} \Big[\frac{(M x+m) k_\perp}{x^2}-\frac{k_\perp (M x+m)}{x^2}- \frac{(1-x) q_\perp (M x+m)}{x^2}\Big]\nonumber\\
&=& 2 M  \frac{g^2}{16 \pi^3} \int dx (M x+m) x^3 (1-x)^4 I_1,
\end{eqnarray}
and
\begin{eqnarray}
B_{\rm b}(q^2)
&=&{-2 M\over (q^1-{\mathrm i}q^2)}
\int\frac{{\mathrm d}^2 {\vec k}_{\perp} {\mathrm d} x }{16 \pi^3}\ (1-x)\
\nonumber\\
&&\qquad\qquad\times\
\Big[\psi^{\uparrow\ *}_{+\frac{1}{2}\, }(x,{\vec k'}_{\perp})
\psi^{\downarrow}_{+\frac{1}{2}\, }(x,{\vec k}_{\perp})
+\psi^{\uparrow\ *}_{-\frac{1}{2}\, }(x,{\vec k'}_{\perp})
\psi^{\downarrow}_{-\frac{1}{2}\,}(x,{\vec k}_{\perp})
\Big]
\nonumber\\
&=& - \frac{2 M}{q_\perp} \frac{g^2}{16 \pi^3} \int\frac{d^2 k_\perp dx}{16 \pi^3} \frac{x^4 (1-x)^4}{L_1^2 L_3^2} \Big[\frac{(M x+m) k_\perp}{x^2}-\frac{k_\perp (M x+m)}{x^2}+ \frac{x q_\perp (M x+m)}{x^2}\Big]\nonumber\\
&=& -2 M \frac{g^2}{16 \pi^3} \int dx (M x+m) x^3 (1-x)^4 I_2.
\end{eqnarray}

Using Eqs. (\ref{diracnpauli}), (\ref{sn1}) and (\ref{sn1a}), the Pauli form factor obtained from the spin-flip matrix element of the $J^+$ current in this case is expressed as
\bea
F_2(q^2)&=& \frac{-2 M}{q^1- i q^2} \left<\Psi^{\uparrow}(P^+,{\vec P_\perp}={\vec q_\perp})
|
\Psi^{\downarrow}(P^+,{\vec P_\perp}={\vec
0_\perp})\right>\nonumber\\
&=& \frac{-2 M}{q^1- i q^2} \int\frac{{\mathrm d}^2 {\vec k}_{\perp} {\mathrm d} x }{16 \pi^3}\ x
\Big[\psi^{\uparrow\ *}_{+\frac{1}{2}}(x,{\vec
k'}_{\perp})
\psi^{\downarrow}_{+\frac{1}{2}}(x,{\vec k}_{\perp})
+\nonumber\\
&& \psi^{\uparrow\ *}_{-\frac{1}{2}}(x,{\vec k'}_{\perp})
\psi^{\downarrow}_{-\frac{1}{2}}(x,{\vec k}_{\perp})
\Big]\nonumber\\
&=& 2 M \frac{g^2}{16 \pi^3} \int dx \left(M+\frac{m}{x}\right)x^3 (1-x)^4 I_1.
\eea
In Fig. \ref{pauliformfactor}(b), we present the Pauli form factor as a function of $q^2$ in the Yukawa theory. On comparing the plots of $F_2(q^2)$ in the QED model and Yukawa theory, we find that their behaviour is almost the same. This is due to the fact that the Pauli form factor is completely dependent on the total momentum transfer in the process and the form factor has connection with the $x$ moments of chiral conserving and chiral flip form factors which appear in deep virtual compton scattering. However, in the limit of zero momentum transfer it gives the anomalous magnetic moment in the simulated model. The Pauli form factor for $q_\perp^2 \rightarrow 0$ comes out to be $F_2(0)=233.912$.

The total contribution at non-zero momentum transfer is given by
\begin{eqnarray}
B(q^2)&=&B_f(q^2)+B_b(q^2)\nonumber\\ \
&=& 2 M \frac{g^2}{16 \pi^3} \int dx \left((M x+m) x^3 (1-x)^4 I_1- (M x+m) x^3 (1-x)^4 I_2\right).
\end{eqnarray}
In order to  check whether the contributions from fermion and boson constituent vanishes or not in this case, we take $q_\perp^2 \rightarrow 0$.
At zero momentum transfer we find that
\begin{equation}
B(0)=B_f(0)+B_b(0)=0.
\end{equation}
For numerical calculations we
have taken $M=m=0.51$ MeV, $\lambda=0$\cite{mod}.
Here also we find that the helicity-flip form factor $B_b(q^2)$ of the graviton coupling to the boson at one-loop order in the Yukawa theory comes out to be $B_b(0)=-475.82$. Whereas, the helicity-flip fermion form factor $B_f(q^2)$ of the graviton coupling to the fermion constituent at one-loop order in the Yukawa theory comes out to be $B_f(0)=475.82$. This leads to $B_f(0)+B_b(0)=0$.
In Fig. \ref{yukawa}, we have presented the  gravitational form factors as a function of $q^2$ in the Yukawa theory and in this case also we obtain the same behaviour as in the case of QED model leading to overall vanishing result for gravitomagnetic moment.

\section{Conclusions}
In the present work, we have considered the light-cone Fock state representation of the composite system consisting of fermion state with a mass $M$ and a vector constituent and a neutral scalar constituent with respective masses $m$ and $\lambda$. We have simulated the wavefunction by differentiating the wavefunction w.r.t. bound state mass $M^2$. The fermion and boson contributions to spin-flip matter form factor in QED and Yukawa model have been calculated and it has been found that the anomalous gravitomagnetic moment $B(0)$ vanishes identically for any composite system in both the  models in the limit $q_\perp^2 \rightarrow 0$. This conclusion is in agreement with arguments based on equivalence principle following directly from the Lorentz boost properties of the light-cone Fock representation. We have also studied the Pauli form factor for both models which is obtained from the spin-flip matrix element of the $J^+$ current and when we compare the behaviour of $F_2(q^2)$ in both models, we find that  due to the connection with the $x$ moments of chiral conserving and chiral flip form factors appearing in the deep virtual compton scattering, their behaviour is almost same. This fact can perhaps
can be substantiated by future experiments in DVCS measurements of the spin flip form factors. New experiments aimed
at measuring the anomalous gravitomagnetic moment would also provide
us with valuable information on the hadron structure.

\section{Acknowledgement}
Authors acknowledge helpful discussion with S.J. Brodsky. HD would like to thank Department of Science and Technology, Government of India for financial support.

\end{document}